\documentclass{aastex63}
\usepackage{colordvi}
\usepackage{hyperref}
\input colordvi

\newcommand{\ha}{H$\alpha$}

\begin{document}
\title{Solar Spicules, Filigrees and Solar Wind Switchbacks}

\author[0000-0002-5865-7924]{Jeongwoo Lee, Haimin Wang, Jiasheng Wang}
\affil{Institute for Space Weather Sciences, New Jersey Institute of Technology, University Heights, Newark, NJ 07102-1982, USA}
\affil{Center for Solar-Terrestrial Research, New Jersey Institute of Technology, University Heights, Newark, NJ 07102-1982, USA}
\affil{Big Bear Solar Observatory, New Jersey Institute of Technology, 40386 North Shore Lane, Big Bear City, CA 92314-9672, USA}

\author{Meiqi Wang}
\affil{Institute for Space Weather Sciences, New Jersey Institute of Technology, University Heights, Newark, NJ 07102-1982, USA}
\affil{Center for Solar-Terrestrial Research, New Jersey Institute of Technology, University Heights, Newark, NJ 07102-1982, USA}

\begin{abstract}
Spicules, the smallest observable jet-like dynamic features ubiquitous in the chromosphere, are supposedly an important potential source for small-scale solar wind transients, with supporting evidence yet needed. We studied the high-resolution {\ha} images (0.10$''$) and magnetograms (0.29$''$) from Big Bear Solar Observatory (BBSO) to find that spicules are an ideal candidate for the solar wind magnetic switchbacks detected by the Parker Solar Probe (PSP). It is not that spicules are a miniature of coronal jets, but that they have unique properties not found in other solar candidates in explaining solar origin of switchbacks. (1) The spicules under this study originate from filigrees, all in a single magnetic polarity. Since filigrees are known as footpoints of open fields, the spicule guiding field lines can form a unipolar funnel, which is needed to create an SB patch, a group of fieldlines that switch from one common base polarity to the other polarity. (2) The spicules come in a cluster lined up along a supergranulation boundary, and the simulated waiting times from their spatial intervals exhibit a number distribution continuously decreasing from a few sec to $\sim$30 min, similar to that of switchbacks. (3) From a time-distance map for spicules, we estimate their occurrence rate as 0.55 spicules Mm$^{-2}$ s$^{-1}$, sufficiently high for detection by PSP. In addition the dissimilarity of spicules with coronal jets, including the absence of base brightening and low correlation with EUV emission is briefly discussed.
\end{abstract}


\section{Introduction}

Since the discovery of magnetic switchbacks (SBs) by the NASA's Parker Solar Probe (PSP) mission, their origin, either the sun or in situ, has been much debated by both solar and space physics communities \citep{Raouafi_2023a,Raouafi_2023b}.
Increasing number of SBs with heliodistance reported by \cite{Jagarlamudi_2023} favors the in-situ origin hypothesis \citep{Squire_2020,Ruffolo_2020,Schwadron_2021}. 
The solar origin hypothesis \citep{Fisk_2020,Bale_2021} is also popular due to the finding that SBs are grouped in a patch structure with a supergranulation size  \citep{Bale_2023}. Open magnetic fields stemming from a supergranulation boundary seem to play a role as a funnel for magnetic transients to escape from the sun to the solar wind \citep{Bale_2021} and granulations seem to be related to individual SBs in size \citep{Fargette_2021} . 
Solar origin models also include Alfven waves driven by granulation creating turbulence in the solar wind \citep{Shoda_2021} and footpoint motions between the fast into slow wind at the Sun creating a magnetic connection across solar wind speed shear, which develops into SBs \citep{Schwadron_2021}.
\cite{Mozer_2021} suggested that SBs form in transition regions, consistent with previous model of reconnection of open field and coronal loop \citep{Fisk_2005}.  

Search for solar originated SBs is often targeted for specific type of solar eruptions. Minifilament eruption producing coronal-jets has often been discussed as a possible solar source of SBs \citep{Sterling_2020b,Neugebauer2021}. Minifilament and flux rope eruptions exhibit signature of magnetic reconnection and twisted magnetic fields, but minifilaments are relatively large and not frequent enough to explain numerous and small SBs.   
Plumelets or jetlets have smaller scales ($\sim$10$''$) and are more numerous \citep{Raouafi_2023a}. 
Recent EUV studies suggested, as potential origins for solar wind SBs and microstreams, EUV ejecta from EUV bright points stemming from usually 1--3 bright points separated by a few arcseconds with periods derived from the fluctuating radial velocities in the range of 3--20 minutes \citep{Kumar_2023b}.
It is yet unclear whether they can produce the combination of supergranulation and granulation scales in the SBs.
An extensive search for EUV jets and {\ha} jets targeted for PSP's detection showed that the occurrence rate of solar EUV jets is not sufficiently high to explain the number of SBs \citep{Huang_2023}.

Spicules are in even smaller sub-arcsec scale in width and much more abundant, which are thus an ideal candidate for the correspondingly small-scale transient in the solar wind. They are often classified into two types: the slowly evolving type-I spicules originated by p-mode leakage in the photosphere and the type-II or fast spicules driven by reconnection \citep{Tsiropoula_2012,DePontieu_2004,Sterling_2020a}.
Ideas on their origin was inspired by \cite{Yokoyama_1996} model for X-ray jets and {\ha} surges due to magnetic reconnection, and includes ambipolar diffusion \citep{Sykora_2017}, momentum pulse at footpoints \citep{Mackenzie-Dover_2021}, and the photospheric flux cancellation \citep{Samanta_2019}. 
Yet there is no clear consensus regarding the origin of spicules and 
the early classification of spicules based on their lifetimes
might no longer be the case with the IRIS observations \citep{Periera_2014}. 
Spicules have, however, a few shortcoming as a candidate for SBs source. First, they are not accompanied by impulsive brightening at their bases (called hereafter jet bright brightening or JBP) if they are reconnection-driven jets \citep{Sterling_2016b, Sterling_2020a}.
Another shortcoming is that correlation between spicular activity and EUV and X-ray jets  remains unclear  \citep{Nived_2022,Uritsky_2021}, and that spicules may not be energetic enough to contribute to coronal heating  \citep{Klimchuk_2012} unlike EUV or X-ray jets. Finally they do not seem to be associated with magnetic eruption that is essential for larger scale solar eruptions.

In order to compare solar data with PSP data, a conversion of solar images to time series of spacecraft measurements has to be made. Lee et al. (2022, hereafter, Paper I) introduced a hypothesis  that spatial structure on solar surface can be mapped into space at the height swept by PSP height at a nominal speed and incident angle of PSP. This is just the reverse of projecting the SB time scales back to the solar spatial scales by converting the PSP time to angular distance traversed by PSP \citep{Fargette_2021,Bale_2021,Bale_2023}. Both studies found the large and medium scale of SBs, which turn out to match the sizes of super-granules and granules.  
It is, of course, debatable whether those variations in the time series are entirely from spatial structure in the sun \citep{Bale_2021, Bale_2023} or also from temporal variation \citep{Shi_2022}. 
On the other hand, a statistical study of PSP data found that the waiting times of SBs appear in a power law distribution from a few sec extending to hours \citep{DudokdeWit2020}. Thus the issue of how to connect PSP to the Sun may not stop at finding of one or two typical sizes of individual solar eruption, but how wide range of scales can be produced from one coherent structure on the sun should also be investigated.

The main goal of this paper is to demonstrate that a clustering of small-scale fluctuations exists in the sun like that of SBs in space. 
Additional goals are to address the aforementioned shortcomings of spicules as solar ejection and to identify the driver of spicules.
Essential for this study is high resolution solar data where small-scale structures are fully resolved. We use such high-resolution data obtained with the 1.6 m Goode Solar Telescope \citep[GST;][]{Goode2012} in Big Bear Solar Observatory (BBSO).
Plan of this paper is as follows: we state the problem in \S 2, and describe major characteristics of the high-resolution BBSO data in \S 3. The results of investigation are presented in the two sections: the magnetic and dynamic properties of spicules and filigrees (\S 4), and their relationship with EUV activity (\S 5). We discuss relevance of these results to the solar wind observations in \S 6, and draw conclusions in \S 7.

\begin{figure}[tbh]  
\plotone{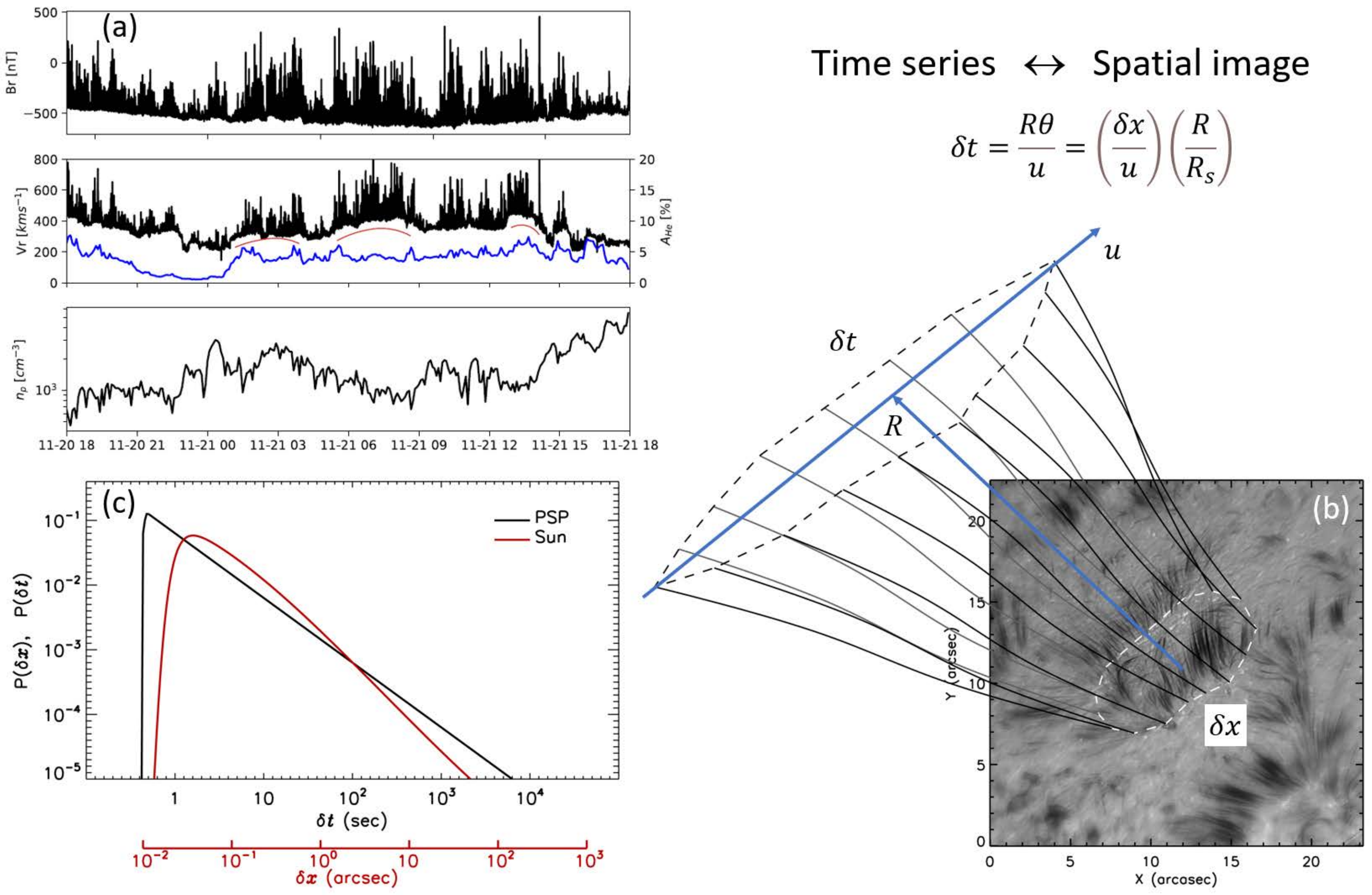}
\caption{In-situ time series data from PSP and solar imaging data fromBBSO/GST. (a) PSP measurements of the solar wind plasma and magnetic field through the November 2021 solar encounter.
(b) A solar image in the {\ha} blue wing from Big Bear Solar Observatory (BBSO) with illustration of field lines originating from a supergranulation boundary to form a magnetic funnel. (c) Probability distributions of PSP time scales and solar spatial scales. The quantities in the conversion relation are denoted in (b, c).}
\label{fig:0}
\end{figure}

\begin{figure}[tbh]  
\plotone{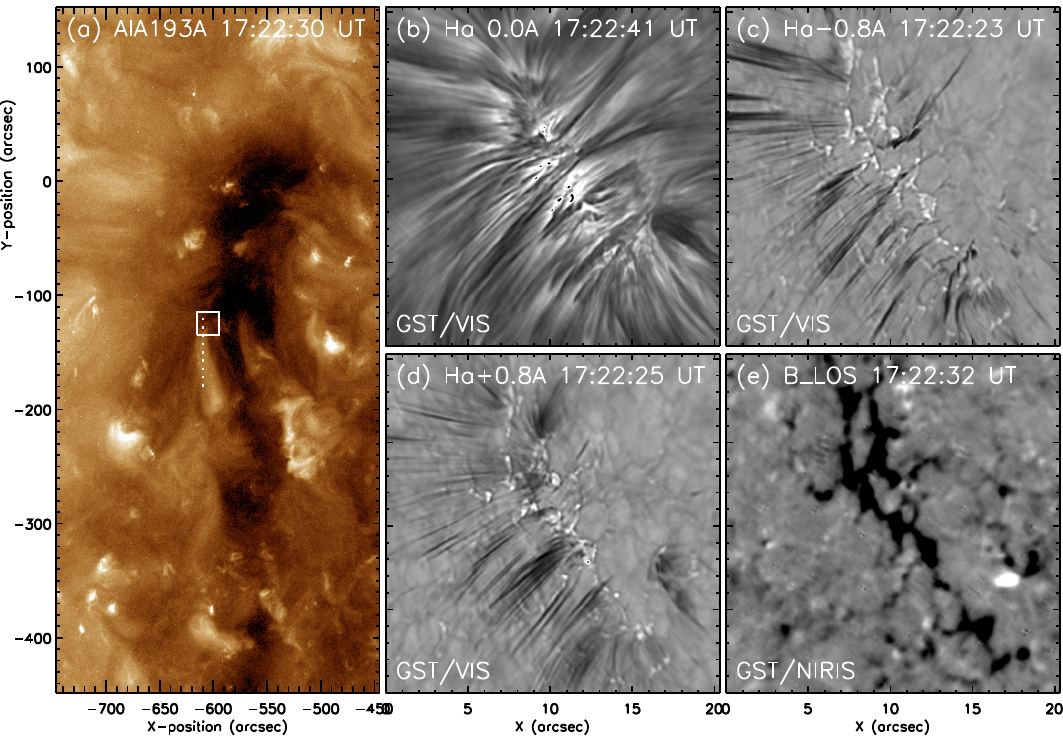}
\caption{SDO/AIA EUV  and GST/VIS {\ha} image observations of a quiet-Sun CH on 2018 July 29.
(a) The AIA 193 {\AA} image shows the location of the disk CH. The white box is the FOV used in the right panels, and the dotted line is a slit used for constructing the time-distance map shown in Figure \ref{fig:10}.
Right panels show BBSO/GST H$\alpha$ images in the H$\alpha$ line center (b) and the off-bands, {\ha}$-$0.8 {\AA} (c) and  {\ha}$+$0.8 {\AA} (d).  A line-of-sight (LOS) magnetogram from the NIRIS is shown in (e). }
\label{fig:1}
\end{figure}

\section{How to Connect the Solar Features to Switchbacks}

Switchbacks are found in the time series of in-situ magnetic field detected by the sweeping PSP, and solar images are obtained from remote sensing, for which comparison between these two sets of data is not straightforward.
Figure \ref{fig:0} illustrates this issue with a sample data from the PSP/FIELD (a), an {\ha} image from BBSO/GST (b), and hypothetical research outputs (c). The PSP data in the arbitrarily selected period shows that the radial magnetic field, $B_r$, solar wind radial velocity, $V_r$, the proton density, and 
the thermal alpha particle abundance ($A_{\rm He}$),
all show the modulation with a patch of SBs with angular size matching that of a typical solar supergranulation \citep{Bale_2021,Bale_2023}. The two red curves in (a) indicate a characteristic variation within a patch, consistent with the diffusion of open field lines within a network as suggested in Paper I. 
If we want to explore individual SBs which must be below the supergranualtion size, we face a couple of observational challenges. No current coronagraph can resolve and trace such small scale solar ejecta. Moreover, higher resolution of a ground-based telescope comes with smaller FOV, and the chance to have a reliable identification of the connectivity between those small FOVs and PSP is extremely low despite great efforts made thus far. 

Suppose we collected all spatial scales from the solar image and want to check them against the time scales detected by PSP (Figure \ref{fig:0}c).
To enable such a comparison, we inevitably make a hypothesis that spatial structure on solar surface can be mapped into space by simple radial expansion at the height of PSP.
Under this hypothesis, timescale $\delta t$ measured by PSP moving at the speed, $u$, at the height of $R$ from the sun with radius $R_s$ is related to the spatial scale, $\delta t$, on the sun as
$\delta t = R\theta / u = (\delta x/u)(R/R_s)$. 
The abscissas in Figure \ref{fig:0}c show $\delta x$ scales lined up with  $\delta t$ scales when PSP was moving at a nominal speed of 500 km s$^{-1}$ at the height of $R= 30R_s$. The often cited supergranulation size, $\delta x\approx 40$ arcsec amounts to $\delta t\approx 30$ min, and a typical mini-filament of size 15 arcsec would result in a time structure wider than 10 min. To address SBs with timescales of a minute or shorter, as is the target of this study, we need to resolve sub-arcsec structures on the sun, at least.
This simple conversion can be altered by many factors including any transport effects, orientation of the SBs relative to the PSP trajectory \citep{Laker2021}, 
and intrinsic temporal evolution of solar sources combined with the finite speed of PSP \citep{Shi_2022}, let alone reliable identification of the connectivity between PSP and the Sun. 

At this point we may ask: if so many factors can alter a certain property of a small-scale solar feature during its propagation, what is the invariant in this Sun-PSP connection?  On the existence of a magnetic funnel with a supergranulation size \citep{Bale_2021, Bale_2023}, the fine structure inside the funnel may retain, at least, a memory of some statistical properties such as number distribution of scales \citep{DudokdeWit2020} and average scales \citep{Fargette_2021, Lee_2022}\label{PaperI}. We do not expect a perfect one-to-one correspondence between $\delta x$ and $\delta t$, and present an example in Figure \ref{fig:0}c, where two distributions somewhat different from each other. Nonetheless solar features, if origin of SBs, may exhibit scales in the range fairly close to that of SBs.
This justifies a study of statistical properties of smallscale solar structures for comparison with SBs dectected by PSP.
In addition, we can utilize high resolution magnetic field data that was unavailable in Paper \hyperref[PaperI]{I}, to hopefully provide further constraints on the possible solar origin of SBs.

\section{Observations}

The data were obtained on 2018 July 29, when the 1.6 m GST/BBSO targeted a quiet-Sun coronal hole boundary (CHB) located about a half solar radius from the disk center at (604$''$ E, 125$''$ S). GST instruments, Visible Imaging Spectrometer (VIS) and Near-infrared Imaging Spectropolarimeter \citep[NIRIS;][]{Cao2012} take advantage of high-order correction by an adaptive optics system with 308 subapertures \citep{Cao2010} and a reconstruction technique for solar speckle interferometric data \citep{Woeger2008} to achieved diffraction-limited resolution under a favorable seeing condition. 
During this observation between 16:34–18:38 UT,  NITIS took high-spatial resolution (0.24$''$) magnetograms and VIS, high resolution (0.10$''$) {\ha} multi-wavelength (11 wavelength points from $-$1{\AA} to $+$1{\AA} of the {\ha} line) images.  Solar Dynamics Observatory Atmospheric Imaging Assembly \citep[SDO AIA;][]{Lemen_2012,Pesnell_2012} and the SDO Helioseismic and Magnetic Imager (HMI) data \citep{Scherrer_2012} aligned with NIRIS magnetogram are used for coalignment of GST {ha} images and AIA EUV images.

 
\subsection{EUV vs. {\ha} emissions}

Figure \ref{fig:1} shows the location of the coronal hole on the SDO/AIA EUV 193 {\AA} image and compares the HMI and the NIRIS magnetograms.
The 193 {\AA} emission is dominated by plasma around 1.5 MK. No large-scale eruptive events occurred near this CH during this time frame.
The white box in (a) is the FOV of the GST/VIS {\ha} images shown in the right panels (b--e) and 
a loop-like structure extends down to this FOV, which is the main EUV structure for this study.
The {\ha} line-center image (b) shows a morphology like parted hair that diverges around the CHB.  The magnetic fields in the CHB have the negative polarity, the main polarity of the CH, as shown in (e).  
In the off-band images (c, d), the {\ha} line opacity drops and the canopy structure becomes faint. As a result we can see granules in the photosphere, and the spicules in the chromosphere as the dark features.
In the EUV image (a), there is no counterpart of these {\ha} fine structures (b--d) in the photosphere and chromosphere.
This exemplifies the difficulty in relating the chromosphere to the corona due to lack of correlation between EUV and {\ha} spicules, as mentioned in Section 1.

\begin{figure}[tbh] 
\plotone{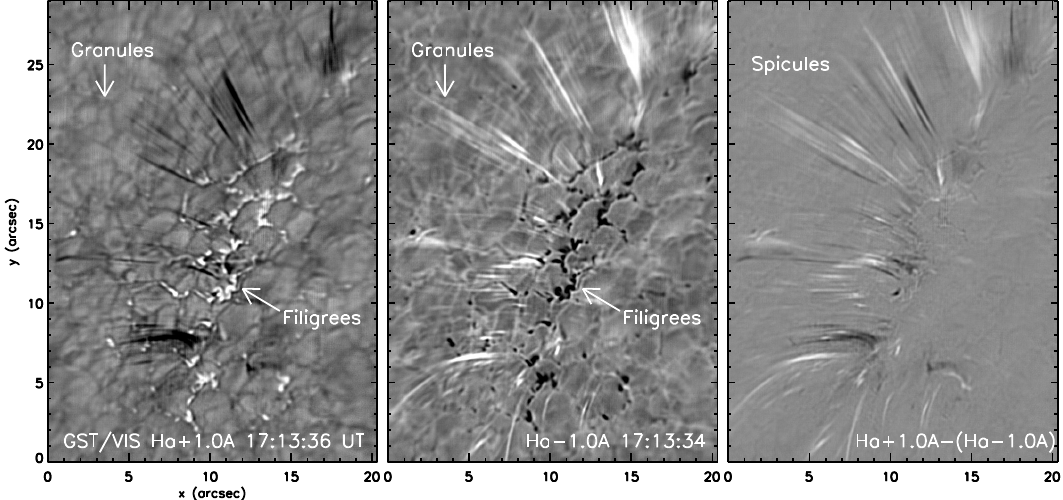}
\caption{Granules, filigrees and spicules in the GST data. (a) The red wing in {\ha}$+$1 {\AA}  (17:13:36 UT), (b) the blue wing in {\ha}$-$1 {\AA} (17:13:34 UT) and (c) a pseudo-Dopplergram constructed from (a,b). Note that (b) is an inverted image.
The granules are more clearly visible in (a) and the spicular activity in (b). The filigrees appear in both wavelengths almost the same, and are subtracted away in the pseudo-Dopplergram (c).}
\label{fig:2}
\end{figure}

\subsection{Spicules, Granules, and Filigrees}

Figure \ref{fig:2} shows spicules and granules against the convective cells in (a) the red wing image, (b) the inverted blue wing image, and (c) a composite of these two wing image intended for a pseudo-Dopplergram.
In the far red wing {\ha}$\pm$1.0 {\AA} of the {\ha} line,  the photosphere is visible without obscured by the chromosphere. It is like a continuum so that the brighter features are actually hotter and the darker is cooler.
Spicules are the only exception to offer opacity in the chromosphere even in the far wings. The red wing images reveal more clearly the convective cell boundary manifested by the darker lane, whereas the blue wing offers some opacity to obscure underneath features, perhaps due to more ejection features. 
Some bright lanes lying along the network boundary are visible in contrast with the ordinary network boundary which appears darker as involved with sinking gas  cooler than the uprising gas in the center of convective cells.
Such bright network boundary in continuum is often referred to as filigrees \citep{Dunn_1973} and the crinkles within a filigree are called magnetic bright points (MBP) or network bright points (NBP). We keep using the term, filigree, because the far wing {\ha}$\pm$1.0 {\AA} images are practically close to the continuum. We avoid using both MBPs and NBPs, because they are sometimes used for X-ray bright points as well. 
Filigrees are believed to the footpoints of open magnetic fields where materials escape out so that the opacity drops to reveal the top layer of convection and they look bright \citep{Dunn_1973,Wilson_1981,Leenaarts_2006,Diercke_2021}. 

In the inverted blue wing {\ha}$-$1.0 {\AA} image (Figure \ref{fig:2}b), spicules and normal convective cell boundary appear white and the filigrees appear in black. 
The granules and filigrees are also visible, not very different from the red wing image.  The pseudo-Dopplergram (c) is constructed by adding (a) and (b) together so that the white (black) spicules are from the blue (red) wing images.
If the spicules in both blue and red wing images coincide each other, we can regard it the line broadening associated with heating. However, they do not coincide each other in position (Figure \ref{fig:2}b, c), and are more clearly seen in (c) as the alternating white and black streaks, which together line up along he magnetic network, forming an outer wall of the network.  
Because of the asymmetric line profiles along spicules, we regard the blue and red wing spicules representing plasma ejection upward and downward, respectively. 
In the pseudo-Dopplergram (Figure 2c), the filigrees are gone as subtracted away, and individual spicules are better visible. Spicules are more or less grouped, which are separated from each other in the scale of one granule or two. The whole length, of course, is about that of a supergranulation so that these spicule distribution possesses two scales corresponding to the medium and large scales of SBs, as reported by \cite{Fargette_2021}.

\begin{figure}[tbh]  
\plotone{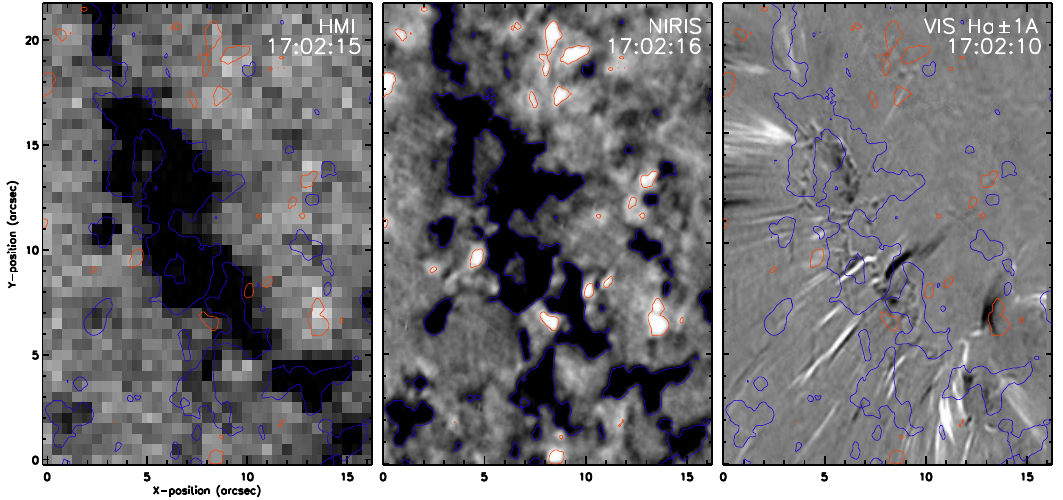}
\caption{Magnetogram and \ha~image from GST in a coronal hole boundary region on 2018 July 29. (a) SDO/HMI LOS magnetogram, (b) BBSO/GST NIRIS magnetogram showing a lot more magnetic elements of minority polarity, and (c) a pseudo-Dopplergram from GST/VIS H$\alpha \pm$1.0 {\AA} wing images. In (c), the white (black) features are upflows (downflows). The NIRIS contours at the levels of $-$40 G (blue) and $+$30 G (red) are overploted in all panels. The lower left corner is positioned at (-601$''$, -148$''$).}
\label{bbso_niris}
\end{figure}

\subsection{Magnetic Fields in the Photosphere}

In Figure \ref{bbso_niris} we compare an HMI LOS magnetogram (a) with the corresponding NIRIS magnetogram (b). The last panel (c) is a pseudo-Dopplergram with the contours of the HMI magnetogram overplotted for comparison. 
The NIRIS magnetogram (Figure \ref{bbso_niris}b) apparently offers a high resolution and sensitivity which allows us to see more bipolar elements (Figure \ref{bbso_niris}a).
Evidently, the NIRIS magnetogram  has a greater sensitivity and higher resolution to reveal more bipolar patches than the HMI magnetogram, which must be useful in detecting small magnetic patches, not only the major polarity patches but smaller and weaker minor polarity patches in studying the interchange reconnection. 
In Figure \ref{bbso_niris}c 
the blue and red contours denote the magnetic fields at the levels of $-$40 G (blue) and $ +$30 G (red), respectively, and show that the magnetic patches in the footpoints of the spicules are in one main polarity, the negative polarity (blue contours).  
Note again that the up and down motions coexist not exactly co-spatial, but together are aligned with along the CHB, meaning that most of them, if not all, are of the negative polarity, the main magnetic polarity in the CHB.
Not only that these spicules have  magnetic bases \citep{Samanta_2019}, but that the bases are unipolar. 
 
\section{Spicules, Filigrees, and Magnetic Polarity}

It is generally accepted that spicules tend to occur in the network boundaries. This is where magnetic fields are not only concentrated and the open and closed field are divided. It is thus worthwhile to check not only magnetic field underneath spicules but its polarity in order to investigate whether spicules are associated with magnetic reconnection or convection or both.  We must also note that spicules under this study refer to those in a coronal hole boundary (CHB), of which magnetic environment may differ from elsewhere.
To define spicules utilize the pseudo-Dopplergram such as Figure 3c, the difference map between H$\alpha+$1.0 {\AA} and H$\alpha-$1.0 {\AA} images. For the filigrees, we use total of the H$\alpha+$1.0 {\AA} and H$\alpha-$1.0 {\AA} images, and define filigrees as regions of intensity contrast defined by $I/\langle I \rangle -1$ being greater than 8\% of the mean background intensity as it best matches the visual appearance. Filigrees and convection cells are visible equally well in the red and blue wings {\ha}$\pm$1.0 {\AA} with a tendency of being slightly better in the red wing, and that spicules are also well visible in both wings, with a tendency that more visible in the blue wing images.

\subsection{Time-dependent Locations of Filigrees}

\begin{figure}[tbh]  
\plotone{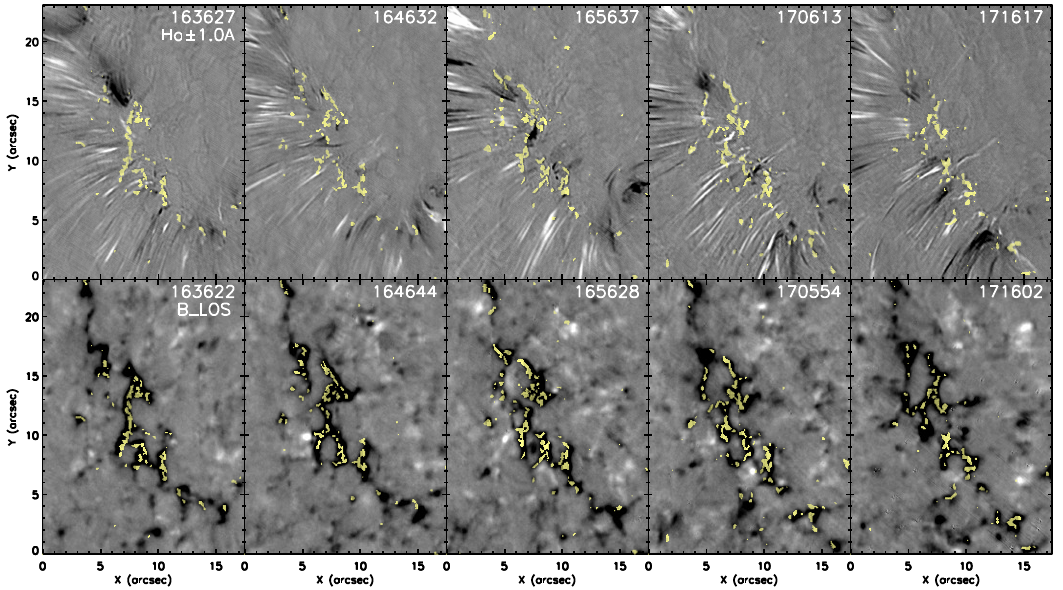}
\caption{Spicular activity and change of filigree distribution. Time sequence of H$\alpha$ wing difference images (top panels) shows ejecting and falling plasma distinguished in white and black colors. Filigrees colored yellow are overplotted here and also on the NIRIS magnetograms (bottom panels). They show shift in position or change in number due to imbalanced birth and decay rates of the filigrees.}
\label{fig:4}
\end{figure}

As filigrees reside in the network boundary and the shape of network may change according to the convection, a question is whether such changes of filigrees are entirely due to changing network fields or also involved with a self-evolution. 
Figure \ref{fig:4} shows the filigrees on the pseudo-Dopplergrams (top) and magnetograms (bottom), and each set of frames separated by about 10 min. The positions of filigrees (marked yellow) determined from the red wing images are copied to the difference images (top) and to the NIRIS LOS magnetograms (bottom).  Each frame is separated by about 10 min apart, and the filigrees do change in this time scale.
Many filigrees are born and decay while the hosting magnetic field changes little. The filigrees either shift in position and increase or decrease in number within a certain magnetic patchy. 
The filigrees form polygons along the network boundary, and evolve from one polygon structure to another with time.  The spicular activity goes along with the convective cells and also tied with the filigrees. We found no apparent causal relationship between spicules and filigrees to surmise that filigrees are not like JBPs. In short, the spicules have the shortest timescale, practically one spicule escapes out of our FOV in a couple of frames (40 s cadence). Next is the filigrees (4-6 min), and the magnetic field in the network boundary ($\sim$10 min). 

\begin{figure}[tbh]  
\plotone{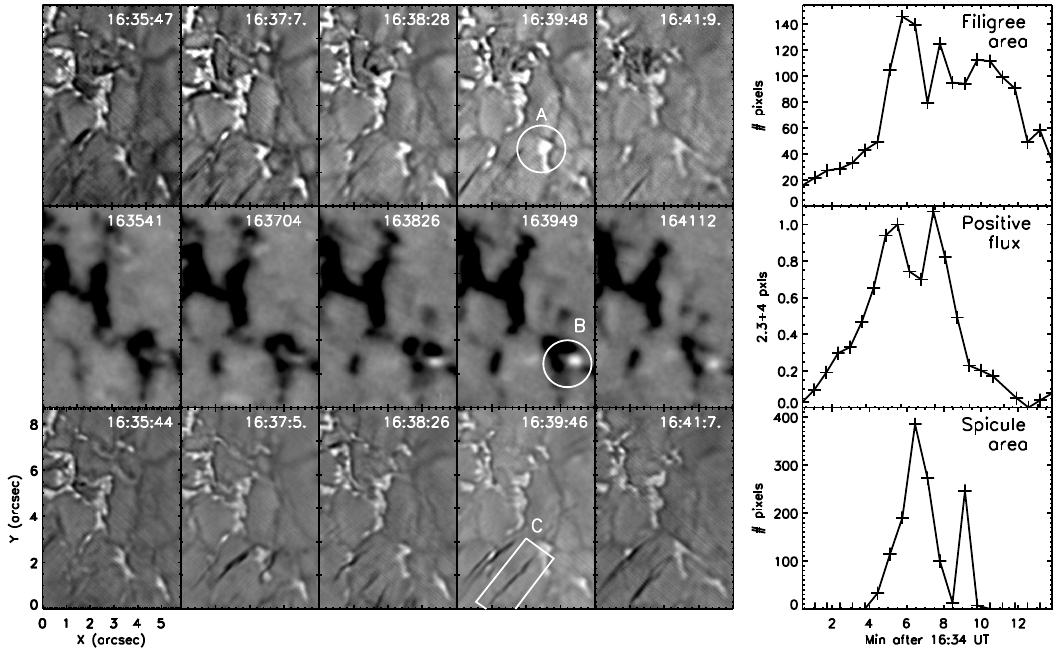}
\caption{Temporal evolution of the spicular activity and associated filigree morphology. (1) In some region,  the filigrees are enhanced, and  ejections occur.
(2) Magnetic field cancellation is found nearby but not at the same locatioon.
(3) Ejections only from filigrees in the main polarity. No ejection from the parasite polarity or PIL.
}
\label{fig:6}
\end{figure}

\subsection{Filigrees as Bases of Spicules}

In
Figure \ref{fig:6}, we check the spicule-filigree relationship by focusing on  a particular spicule occurring near a cancelling bipolar element. The filigree is marked in the red wing image (top panels) and the engaging spicule is marked with a rectangle in the blue wing image (bottom panels).   
Although magnetic cancellation is going on (middle panel), 
a close look finds that the spicule (marked by while arrow in the bottom panels) is coming from the filigree (circle A), the flux cancellation (circle B) is between the minority polarity patch and the nearby main polarity one (both are not filigrees).
The filigree (circle A) is of course in the main polarity and not paired with the other polarity, and is therefore located in a little shifted position from the PIL (circle B), and the spicules is also off the PIL.
In terms of the overall temporal variations of the magnetic flux in the positive polarity and the area of spicules in the $-1.0$ {\AA} image gives an impression that spicule is apparently associated with magnetic flux cancellation. However, in position, the spicule is about an arcsec off from the cancellation bipole, and is connected to the nearby filigree (see  the white box in the bottom panels).  
This result implies that the flux cancellation or reconnection underneath may not be the main driver for spicules.

\begin{figure}[tbh]  
\plotone{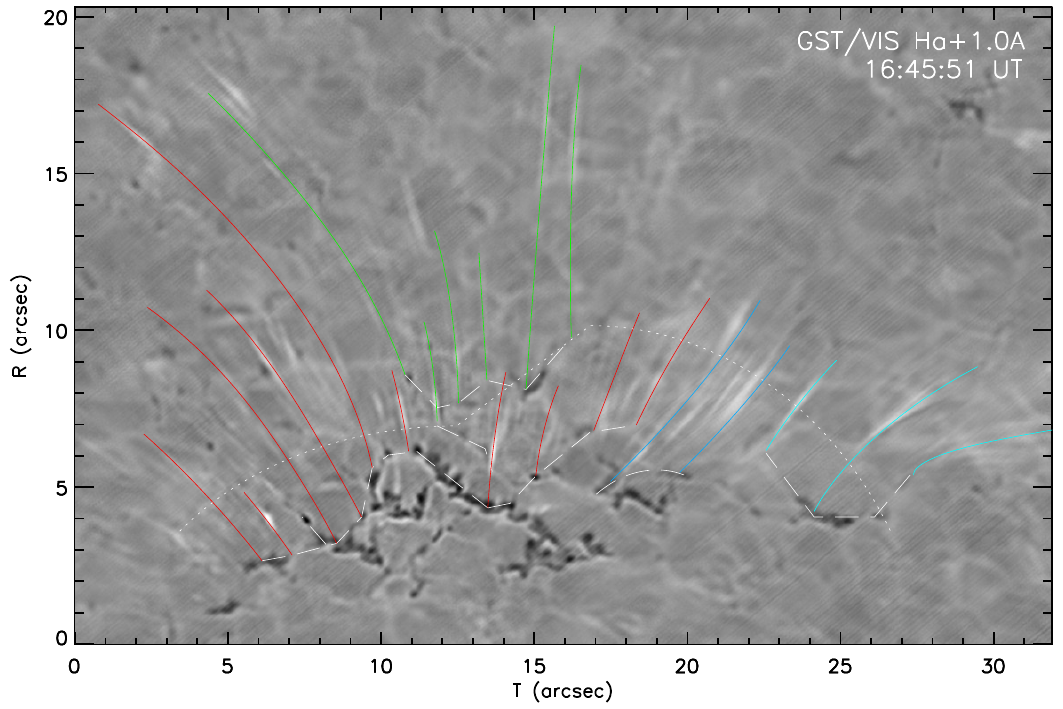}
\caption{Spicules and filigrees.
An inverted {\ha}$-$1.0{\AA} image at 16:36:25 UT from GST/VIS is rotated so that the limb direction is to the $y$-axis. The center of FOV is about (-603$''$, -125$''$).
In this inverted image, the dark features are filigrees, and the white and long features are spicules.
Granule boundaries are visible as white lane of polygons. 
The colored lines delineate the spicules and the dashed lines, relevant filigrees. The dotted line is a slit used for constructing the t-d map in Figure \ref{fig:9}.
Animation shows the inverted {\ha}-1.0{\AA} images (without the guide lines) for total duration of 1 hour 35 min.}
\label{fig:7}
\end{figure}

In Figure \ref{fig:7}, we investigate the spicule-filigree relationship over a wider FOV. The background images is an inverted {\ha}$-$1.0{\AA} image from GST/VIS, where bright features are spicules and convection cell boundaries. 
The image is rotated to the direction where the limb is located toward the positive $y$-axis. 
Only obvious trajectories are marked with random colors, because some of them are too faint. They appear as straight or only slightly curved lines, although twisted ones are occasionally found.   
Some spicules are visible from the footpoints, and some could be detected at greater heights.  
Nonetheless we could identify the footpoints of spicules by either extending the instantaneous trajectory down to the photosphere or trace the motion of spicules in consecutive images back to the photosphere.
At a glance, 
the spicules emanate from the granule boundaries, supporting the magnetic bases of spicules \citep{Samanta_2019}. A more careful examination reveals that spicules are sporadically ejected from only a set of selective filigrees at each time.
The filigrees therefore work like place holders for spicules in the sense that open flux tubes connected to the filigrees would work as a conduit for spicules. 
 
\begin{figure}[tbh]  
\plotone{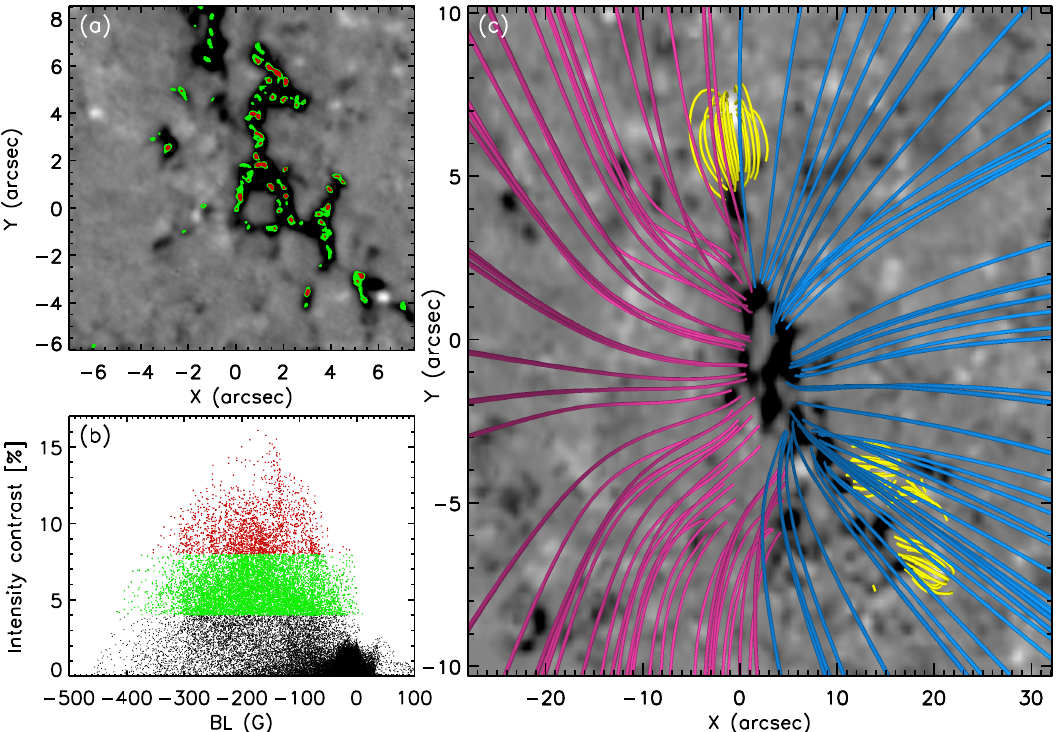}
\caption{Filigrees and open magnetic field lines. (a) Two levels of filigree intensity contrast, 4\% (green) and 8\% (red) of the {\ha} line intensity in the wing images are marked with different colors. (b) Scatter plot of the intensity contrast and magnetic field. (c) The potential field extrapolation show 
the open field lines distinguished by three colors depending on whether directed to the east (purple) and west (blue), and closed field lines (yellow).}
\label{fig:8a}
\end{figure}

In Figure \ref{fig:8a}, we investigate the relation of filigrees with magnetic fields.
By plotting filigrees over the NIRIS magnetogram (Figure \ref{fig:8a}a), we can see that the filigrees appear in not all but a subset of the negative polarity regions. To see whether filigrees have internal structures depending on the field strength, two colors are used for two levels of filigree intensity contrast, 4\% (green) and 8\% (red) of the {\ha} line intensity. The scatter plot in Figure \ref{fig:8a}b shows that the intensity contrast is positively correlated with the field strength to some field strength, after which the correlation turns over. In regions where the intensity contrast is low, both polarities may appear around zero magnetic field, i.e., field-free region. However, above our threshold for filigrees (8\%), no exception is found from the rule that all filigrees are in the negative polarity. 
In Figure  \ref{fig:8a}c, we check this rule again using the potential field extrapolation. 
The purple and blue lines represent the open fields diverging toward the east and the west, respectively, which resembles the combed-hair-like structure seen in the {\ha} centerline (Figure \ref{fig:1}b). The yellow lines are closed fields underneath the canopy, which takes the form of anemone structure or embedded bipole \citep[e.g.][]{Kumar_2023b}. Contray to our expectation, embedded bipoles are not well correlated with spicules.
Filigrees are definitely connected to open fields in negative polarity. This result may look trivial because most of the negative flux patches are connected to open fields, as expected for a coronal hole region with the negative polarity. 
It is, however, notable that the negative flux patches, if located near the positive flux patches, do not have filigrees. Since bipolar pairs are likely of closed fields, the absence of filigrees on them also obeys the rule that filigrees are tied to open fields. It is important for the context of this study that all filigrees are connected to the open fields sharing the same magnetic polarity, and thus spicules are moving along the open fields in a single polarity.  

\subsection{Time-Distance Map for Spicules}

\begin{figure}[tbh]  
\plotone{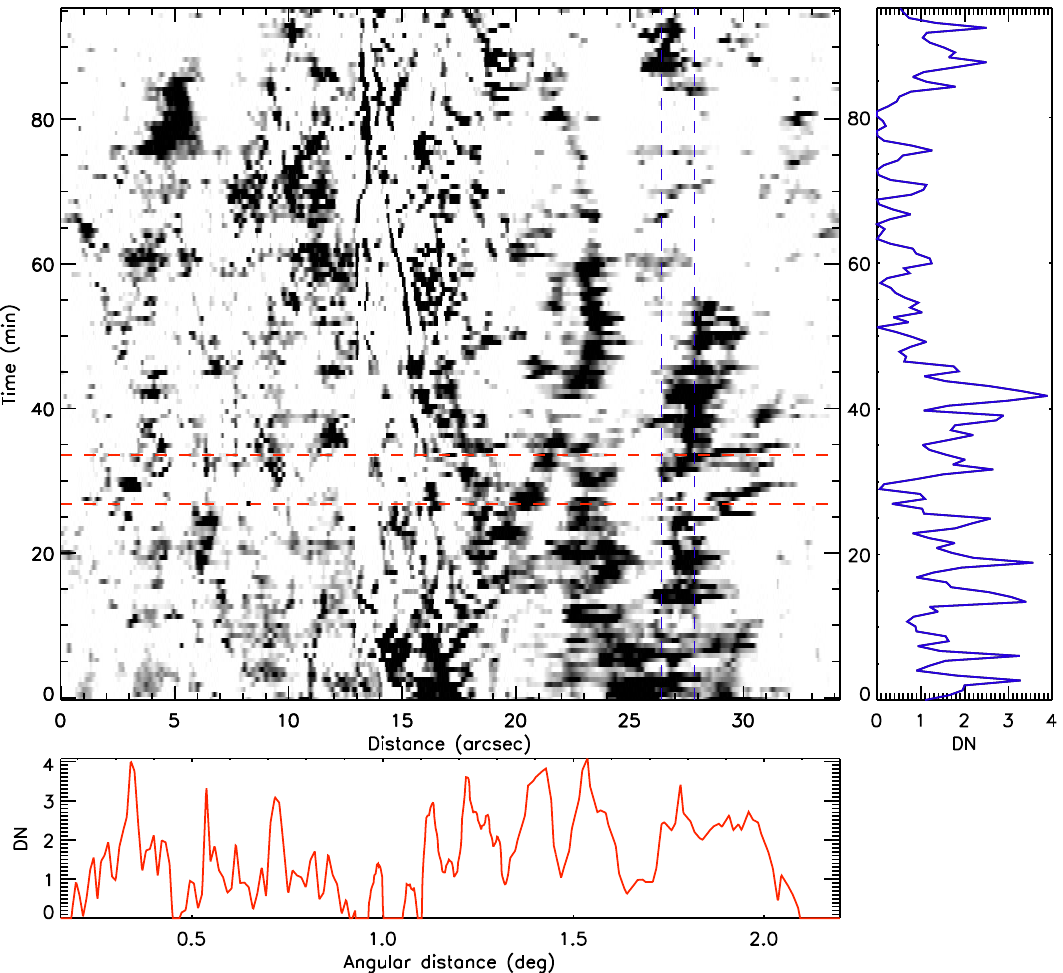}
\caption{T-d map for spicules (center) and readouts along the position axis (right) and along time axis (bottom). The time axis is in units of minutes. The positions are given in both distance (arcsec) and angular distance (degree).}
\label{fig:9}
\end{figure}

Tracing highly transient spicules must be challenging, and we intend to utilize the time-distance map. However, it turns out that spicule motion is hardly detectable at our 41s cadence for the reasons explained below. 
The time-distance map in
Figure \ref{fig:9} is constructed using the set of slits denoted in 
the top panel displayed together with two readouts along the time and the distance axes (red and blue dashed lines).
In the map, we surpressed bright features such filigrees and convection cells, to focus on the spicules which appear as dark features.
Some of them appear to continue at consecutive times at a fixed position, and some others slip sideway, but spicules detected over consecutive frames are rare in our data. Therefore we regard most of the dark features in the time-distance as simply instantaneous appearance of spicules in every single frame, and utilize this time distance map for no more than counting events.
The readout shown in the right panel is a time profile of the sign-reversed DN, which would be a proxy for either enhanced density or temperature. It tends to show 3--10 min quasi-periodicity, but this apparent periodicity arises due to intermittent birth of spicules at multiple locations rather than a coherent oscillatory motion persisting at a fixed spicule. Nonetheless they may somehow contribute to the 3--10 quasi-periodic EUV fluctuations in the corona, which are claimed to be solar sources of micro-streams \citep{Kumar_2023b}.
 
On the other hand, the spatial distribution of spicules read out along the red band (bottom panel) takes after typical SB structure (see, for example, Figure \ref{fig:0}a). The distance is now given in units of degree for comparison with a satellite data independent of its height. 
The typical scale of this flux tube distribution is found to be 8--18 arcsec apart or 0.05--0.11 degrees, and the whole length of the region under consideration is 30--35 arcsec, or $\sim$2 degrees. As a matter of fact, PSP moves at a finite speed, and would scan this t-d map at a slant angle to have a spatial structure longer tha shown in this bottom panel. In any case, these scales agree to the known scale of individual SBs and that of an SB patch \citep{Fargette_2021}.
The spatial profile rapidly changes with time, and practically only a very few spicule structure is repeating in the adjacent frames. Since PSP is moving at a finite speed, it will sample the spatial structure at different times in a path. Further depending on the incident angle, the path length of PSP through one patch may increase more compared with what is shown here. While a funnel (an SB patch) is more or less a spatial structure, the ejecta scanned by a flying spacecraft forming fine-scale structures inside a patch may include temporal variation. The number of SBs within a patch may be invariant, though.  Typically $\sim$20 SBs should be inside one patch.


The t-d map of spicules in this case works only as a marker of spicules in space rather than tracing continuous motion over many frames. We would rather determine the spatial structure at each time, and add their information over  all time bins of the observation to put into the number distribution. 
(a) We fit each peak over some threshold with a Gaussian. We remember two quantities: the location and the width of the Gaussian fits.
For compare the result with the time distribution \citep{DudokdeWit2020}
Repeat over many times permits us to acquire enough number of statistical distribution. 
As a result, (b) the inter-distance distribution (when converted to time distribution assuming height of 30$R_\odot$ and PSP's speed of 500 km s$^{-1}$. 
We 
convert the spatial scales into time scales as illustrated in Figure \ref{fig:0}.
The time distribution has a monotonically decreasing function from 1 se to 1.8$\times 10^3$ s, with lower end similar to the switchbacks timescales \citep[e.g.][]{DudokdeWit2020}. The longer waiting time is missing because our FOV is limited. Other than that it quite overlaps with significant portion of the waiting time distribution, although the distribution may change depending on the speed of PSP and the incident angle to the structure.
(c) The diameter distribution has a peak at 0.16 arcsec and decreases both ways. Missing larger scale is of course spicules cannot larger than a certain size, whereas The peak corresponds to 7 s  (116 km). This has to be compared with typical width of the SBs.

\begin{figure}[tbh]  
\plotone{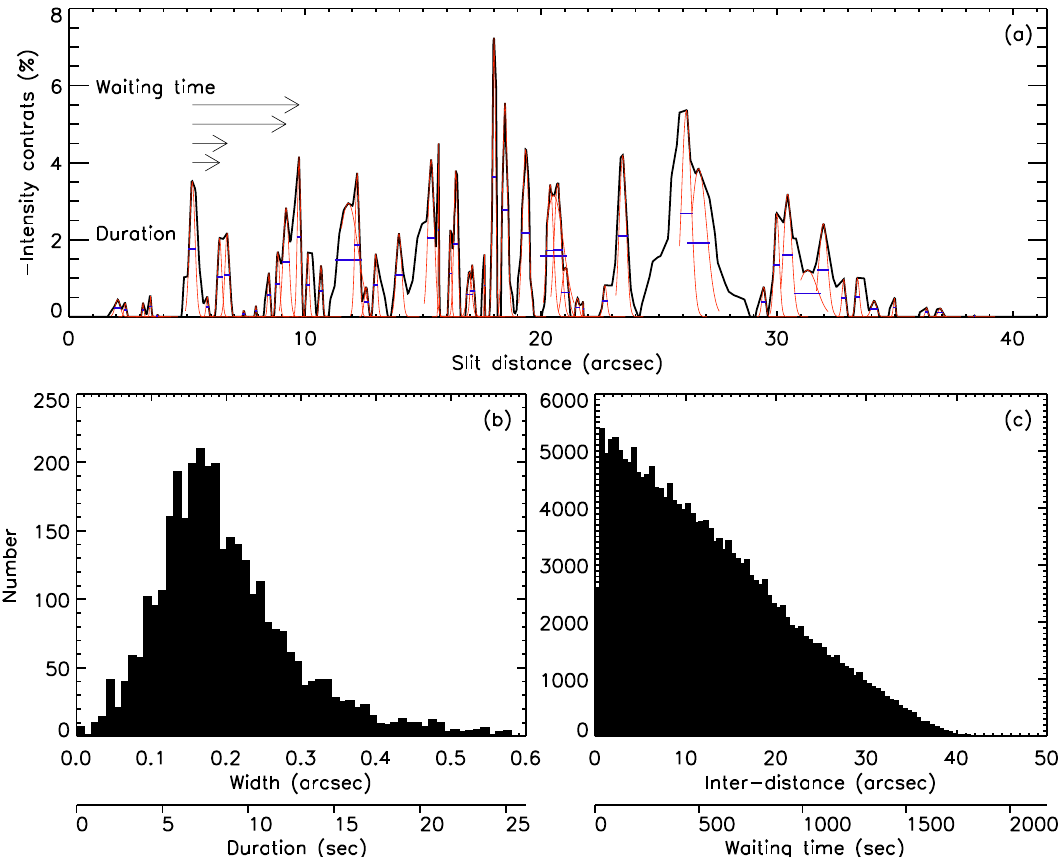}
\caption{Statistical properties of  spicules. (a) 1D plots of t-d plot read out gaussian fit is made around the peak of intensity contrast greater than threshold. (b) Widths of spicules and (c) inter-distances among spicules. Spatial scale in the sun is converted to angular distance and time intervals to be measured by a spacecraft at 30$R_\odot$.}
\label{fig:9a}
\end{figure}

\section{Spicules and EUV Emission}

\begin{figure}[tbh]  
\plotone{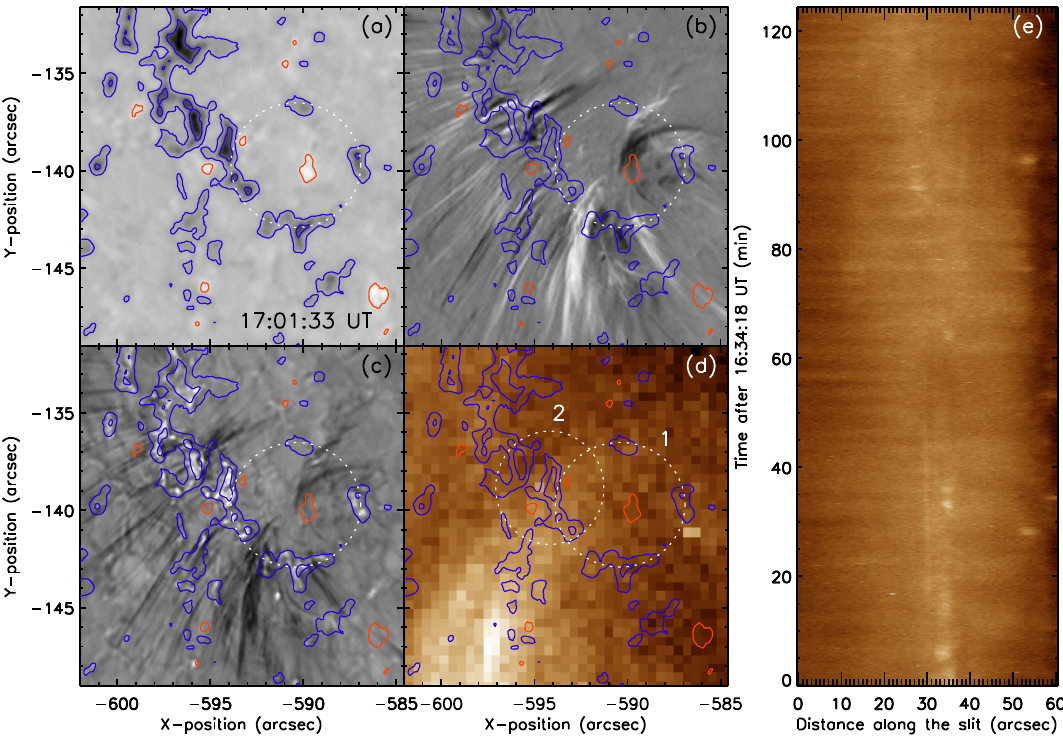}
\caption{H$\alpha$ spicular activity and EUV transient brightenings.
(a) NIRIS magnetogram with contours at the level of $-$[50, 200] G (blue) and [50] G (red). These contours are reproduced in other panels.  (b) GST/VIS H$\alpha$ blue-red wing difference image and (c) total of the blue and red wing images at 17:01:33 UT.
(d) SDO/AIA 193 {\AA} image with the magnetic field contours.
(e) Time-distance map of the EUV intensity along the slit denoted in Figure \ref{fig:1}a.}
\label{fig:10}
\end{figure}

In general, spicules and EUV emission do not correlate well with each other \citep[e.g.][]{Nived_2022}, which yields an impression that they are not driven by reconnection. We nonetheless look for a possible relation between them by comparing spicule activity and magnetic property in the regions with and without EUV brightness.
Figure \ref{fig:10} shows (a) NIRIS magnetogram, (b) pesudo-Dopplergram, (c) total filtergram, (d) EUV image, and (e) t-d map. The t-d map constructed from the slit denoted in Figure \ref{fig:1}a shows that occasional flickering of EUV brighteness occurs inside this EUV loop-like structure.   
Those EUV flickering patches have strong spicules coexistent within the loop-like structure (of a granular size), which may therefore be related to each other. 
On the NIRIS magnetogram (a) the region marked with a dotted circle 1 appears to be a good candidate for the source of the EUV loop structure. It has a parasite polarity patch surrounded by the major polarity fields, also called an embedded magnetic bipole, so that it can form a null structure in the pseudo streamer.
However, the {\ha} pseudo-Dopplergram (b) and total intensity (c) shows that the ejections from its center is oriented toward the west, where the EUV 193 {\AA} emission is rather dim (d). The arc-shaped EUV loop structure are directed southward rather than eastward.  
In terms of position, another candidate region, circle 2, is closer to the footpoints of the EUV  loop-like structure, in which case the tiny parasite patches in the minor polarity (red contours) around the main polarity lanes could have caused the EUV brightness flickering. 

\begin{figure}[tbh]  
\plotone{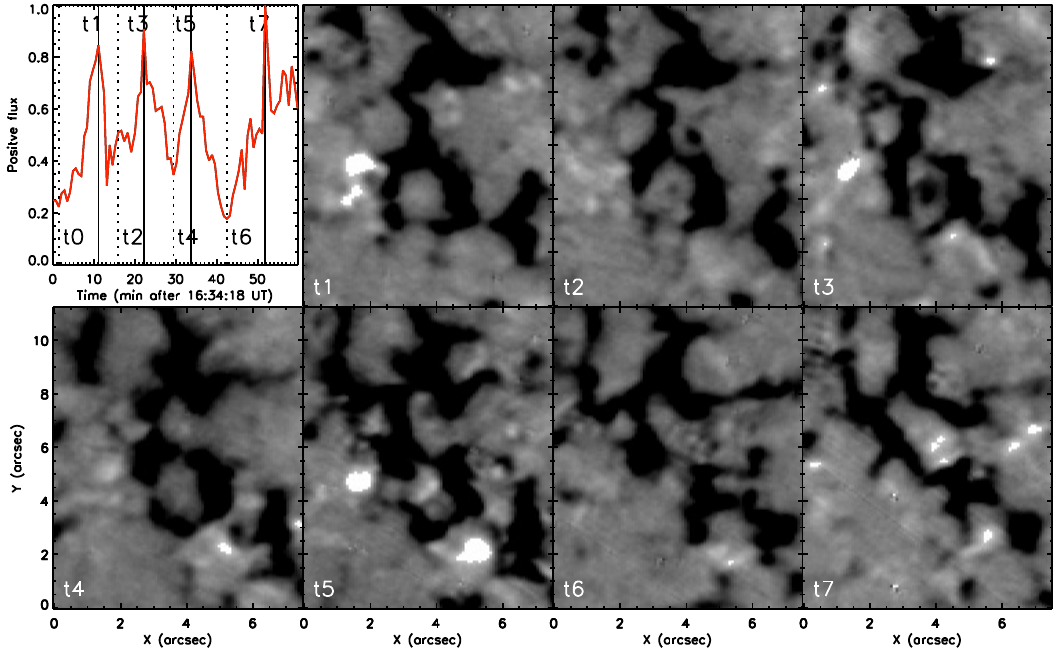}
\caption{Temporal variations of the minor polarity flux and selective NIRIS magnetograms. The quasi-periodicity of 8--10 min in the flux time profile of the minority patch (positive polarity) and the magnetogram is clearly visible only in  region D, which is associated with the EUV brightness.}
\label{fig:12}
\end{figure}

We investigated the magnetograms for a repeating pattern of the minor polarity flux in the region of the EUV activity.
In Figure  \ref{fig:12} 
the first panel shows the flux time profiles of the minor polarity field in the FOV, and the rest panels show selected magnetograms.
The minor polarity patches keep coming up and disappearing,
so that the time profile of the positive flux shows a quasi-periodic variation with about 8--10 min interval. 
This quasi-period is close to a turnover time of a convection cell, and also falls in  the longer side of the timescales from the fluctuating radial velocities in EUV emissions approximately 3--20 min \citep{Kumar_2023b}. The magnetic fluctuation may perturb the field lines to cause reconnection and subsequently the EUV brightness. It could also cause motions of footpoints motions to have MHD waves generate and propagate out.
This quasi-periodic variation of the minor polarity patch is detected only in this region for a limited time period, while spicules occur elsewhere. This is the region where the extrapolatied field lines run along the loop-like EUV bright structure, and the spicular activity is correlated with the EUV emission. The result is in favor of the selective correlation between {\ha} ejection and EUV emission found by \cite{Nived_2022}.

\section{Discussion}

We have investigated whether spicules qualify as a candidate for the  solar seed of SBs.
For this goal, it essential to resolve the following two issues: why spicules lack of the usual reconnection features found for the coronal jets, and whether they are involved with interchange reconnection.
We obtained both the negative and the partially positive results and discuss them in the below.
 
\subsection{Missing JBPs and EUV Counterparts}

We were unable to find obvious signatures for magnetic reconnection in the immediate vicinity of spicules. 
Magnetic cancellation frequently occurs but not directly underneath spicules. 
Twist and rotation of the spicules are visible only occasionally, and most spicules are straight or only weakly curved (Figure \ref{fig:7}). From the poor temporal correlation between spicules and filigrees, we suggest that filigrees underlying spicules are not like JBPs.   
Correlation of spicules with EUV flux is also low, and we suggest that spicules are not a miniature of coronal jets.
It is however debatable whether the absence of such magnetic reconnection signatures should preclude spicules as a candidate for SBs.  
We must note that
a twisted structure, although has been of interest for the flux rope candidate, does not really explain the SBs structure, because SBs are essentially kinked magnetic structures. Many models simply assumed that a kinked structure will form as a result of interchange reconnection. However, a kinked structure will immediately flatten out in the corona due to dominance of magnetic force over plasma pressure. 
\cite{Owens_2018} resolved this issue by placing the reconnection point higher up, where such a kink can survive against flattening. Possibility of an extended chromosphere has been proposed by \cite{Moore_2023} in another context.
In this case, the JBPs may be too faint to be detected in the photosphere. Furthermore JBPs are expected for internal reconnection, and not for outer reconnection, i.e., reconnection occurring above the ejecta.
The different behaviors of spicules with those of EUV activity can be reconciled if they are like cool and hot jets in the BP models.
Cool jets in the BP model \citep{Scott_2022} are driven by the rarefaction waves in high beta plasma, in which case neither twisted field lines nor JBPs are necessary. 
They can just be ejected along the open fields rooted in filigree, which work as a conduit for chromospheric ejecta. 
Even though passively responding to the magnetic reconnection occurred high above,  spicules are still an important manifestation of the interchange reconnection.

\begin{figure}[tbh]  
\plotone{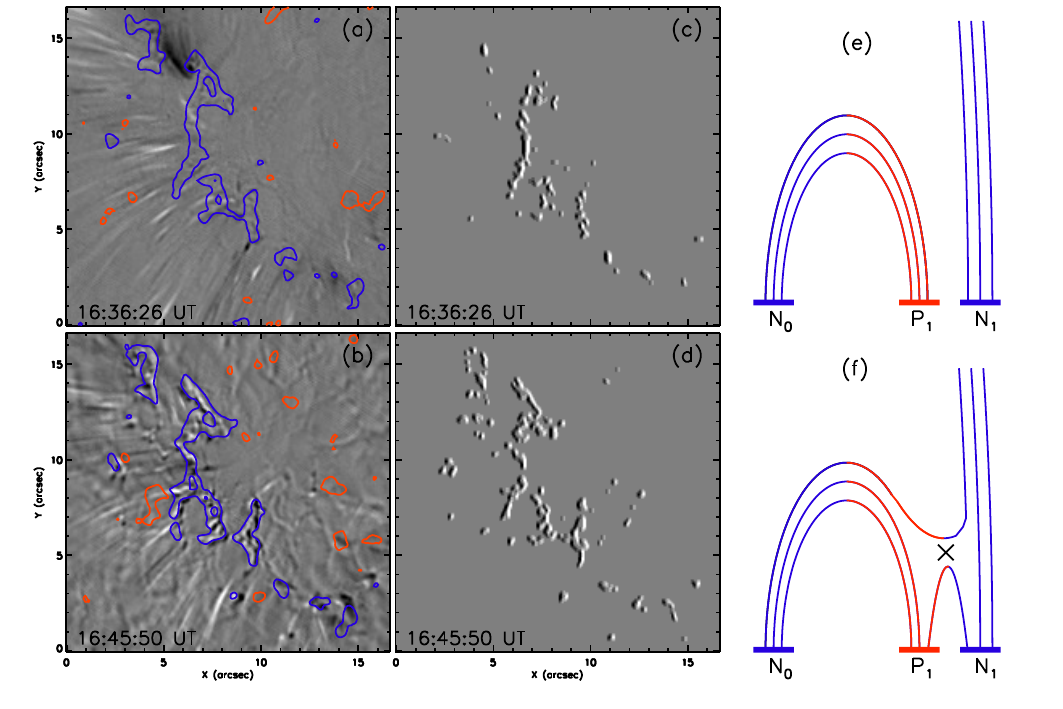}
\caption{
Evolution of filigrees and inferred magnetic field in the corona.
Magnetic field (contours) are plotted over the pseudo-Dopplergram at two times separated by about 10 min (a, b).
Distribution of filigrees at the corresponding times are displayed using the EMBOSS function (c, d) to show their positional shift and birth or decay moore clearly.
Schematic cartoons illustrate how such changes in filigrees may occur due to the change of the open-closed field boundary under interchange reconnection (e, f).}
\label{fig:13}
\end{figure}

\subsection{Frequent Interchange Reconnection}

Interchange reconnection has been the most often discussed mechanism for producing SBs \citep{Zank_2020,Liang2021,Drake2021,Telloni2022,Bale_2023}. It is therefore necessary to check if spicules are also driven by interchange reconnection.
In the present case, the filigrees are in the single magnetic polarity, meaning that the open fluxtubes guiding the spicules are share one polarity, and the other polarity fields must be in the form of closed loops. 
This circumstance makes it obvious that interchange reconnection must be the most plausible mechanism for the spicules. 
A more meaningful question would be how many events of interchange reconnection are occurring in this region.

In this regard, we relate the positional shift and the birth/decay of filigrees to the connectivity change as a consequence of the interchange reconnectionin the open-closed field boundary. In Figure \ref{fig:13}
The pesudo-Dopplergrams (a,b) made using H$\alpha \pm$0.8 {\AA} at two different times with the NIRIS longitudinal magnetograms as contours. While the magnetic field only little changed, the spicules and filigrees significantly changed in both position and number. The change in the filigrees are more  emphasized using emboss function in IDL for highlighting only bright features above the threshold intensity contrast of 8\% (c,d).  
In the cartoons (e, f), the system has  a bipolar element, $P_1$--$N_1$  approaching to $N_0$ (e). At this time  only one filigree can exist above $N_0$ as it is the only open field, while $P_1$--$N_1$  are footpoints of a  closed field not to have any filigrees. 
After the interchange reconnection (f), $N_1$ becomes another open field to host another filigree. The former filigree, $N_0$, may either disappear or remain in an altered intensity. 
If $N_0$ disappears, this process will be viewed as a positional shift of the filigree; otherwise we will see an increase of filigrees in number.  
Either way, the positive flux, $P_1$, can never be open, and the filigrees are mostly on the negative flux patch. 
Under this circumstance, the change of filigrees in position and/or number 
can be a manifestation of interchange reconnection higher up in the corona.
 In the sense that a newly appearing filigree manifests a transformation of a closed-to-open fluxtube, the increasing number of filigrees may work like that of alpha particle abundance, $A_{\rm He}$.

What is important for understanding the SB structure is not a single event but a series of interchange reconnection occurring in the spatial intervals of about granule size.
We witness in our data that filigrees form a set of polygons connected over a longer distance (about 30 Mm) than in other regions, each one forms a honey bee nest like structure in the scale of one or two granules (about 1 Mm), which manifests the SB structure on the solar surface. 
This result may complement the EUV study, where the change of the complexity along the CHB is taken as evidence for interchange reconnection \citep{Mason_2022}.
It would be more useful to count the number of spicules that  occurred for a given duration and an area, since filigrees do not have one to one relation with spicules. We counted 3,459 spicules on this t-d map over total observation time of 95 sec sampled over the area of 66 Mm$^2$. This gives 0.55 spicules Mm$^{-2}$ s$^{-1}$. Namely, one spicule is ejected over an area of two granules every sec, which is a very high occurrence rate compared with any other solar ejecta.

\begin{figure}[tbh]  
\plotone{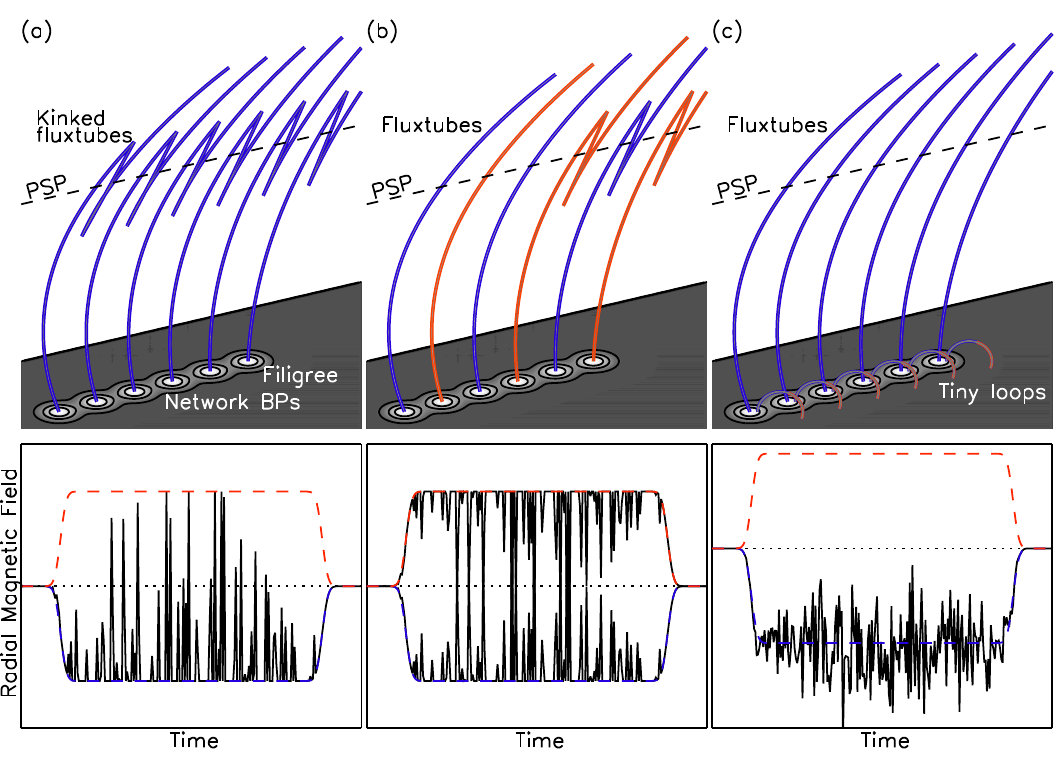}
\caption{Hypothetical measurement of the radial magnetic fields by a satellite (bottom panels) by a satellite expected under  three magnetic configurations (top panels): (a) unipolar open flux tubes lined up along the network  boundary and develop kin high in the corona, (b) multipolar open flux tubes kinked or not, and (c)  flux ropes formed in the low atmosphere to be embedded within the open fields.
Dotted lines are zero magnetic field, taken as the background field in the solar wind.
}
\label{fig:14}
\end{figure} 

\subsection{A Constraint on Magnetic Polarity}



The aforementioned single polarity of the filigrees appears to be be a crucial factor in relating solar features to SBs,  because field lines within an SB patch should be unipolar below the point of kink, after which they switch to the other polarity and come back to the original polarity. 
In Figure \ref{fig:14}, we plot three magnetic configurations (upper panels) and the expectant magnetic field measured in space by hypothetical PSP (lower panels).
When a group of open flux tubes rooted in the filigree of a single polarity make a kink in space, the PSP will find the sign-reversing fields, which are deflected toward the other polarity from the baseline magnetic field as looking like the SB structures detected by PSP (Figure \ref{fig:14}a).
These fluxtubes can interact with neighboring fields only through interchange reconnection, because the other minor polarity has to be in the form of closed field. 

If filigrees are in the mixed polarities as in Figure \ref{fig:14}(b) the baseline field itself is alternating in signs. They also look like SBs, but do not satisfy additional conditions like electron pitch angle distribution. The kink will only add more complexity in alternating in signs, which does not make them qualify SBs. In this case, reconnection may occur between two opposite directed open flux tubes and flux cancellation may also occur.  None of these help producing SB properties.
If twisted fields or flux ropes are added as in Figure \ref{fig:14}(c) it will result in field deflection in both directions around the average base field (dashed line).
The scenarios shown in in Figure \ref{fig:14}(b, c)  cannot make the SB structure. 
The most natural explanation is available when a group of unipolar flux tubes is kinked higher in the corona as depicted in Figure \ref{fig:14}(a).

To complete this argument we should be able to provide a mechanism for transforming spicules into the SB structure. Since the spicules move in more or less straight trajectories within our FOV, we presume that the fluxtubes guiding must develop a kink in the higher corona. This hypothesis  is supported by \cite{Owens_2018} model in which a kinked structure resulting from interchange reconnection may be preserved into space if the reconnection occur in the sufficiently high corona. 
Other models in line with this hypothesis include formation of the kinked field lines by shear motion \citep{Schwadron_2021} and so-called N-wave mechanism that an outward-propagating shock–rarefaction system overtakes the leading shock to form a shock–rarefaction–shock triplet to become kinked structure in space \citep{Scott_2022}.

\section{Conclusion}
We have used the high resolution magnetograms from GST/NIRIS  and  {\ha} filtergrams from GST/VIS to investigate the magnetic and statistical properties of spicules from a disk CHB to find them a strong candidate for solar seed of SBs. 
It is not that we found those spicules a miniature of well-known coronal eruptive phenomena, but that spicules's own properties as chromospheric ejecta come out favorable for explaining SBs. 
Of these, three particular properties appear essential as a solar source candidate for SBs: (1) they originate from single polarity filigree bases, (2) are clustered with a wide range of spatial scales, and (3) occur at a sufficiently high rate.
Why these are important factors for solar originated SBs are as follows: 

\begin{enumerate}

\item
To reproduec the polarity of SBs, the field lines before entering SB region should be in a single polarity, later switching to the other polarity by deflection, and back to the original polarity. This leads to the not yet fully discussed unipolar condition, which will be met by the flux tubes rooted in the filigrees in a single polarity, but neither by flux tubes in mixed polarities nor by erupting flux ropes. 

\item
Interchange reconnection, the most preferred mechanism for SBs, is obvious in this magnetic setting consisting of the unipolar filigrees and minor polarity fields around in the closed form only. More importantly, the numerous filigree's shift or increase/decrease in number indicates correspondingly high rate of interchange reconnection along the CHB, analogous with the rapid time-dependent change of CHB shape in EUV images \citep{Mason_2022}.
We also demonstrate that the interchange reconnection should occur higher in the corona to preserve the kinked shape suitable for the SB structure. \citep[cf.][]{Owens_2018,Moore_2023}.

\item
To explain a patch of SBs, not just 
a single instance of interchange reconnection but a series of interchange reconnection must occur in the anngular spatial  scales matching those of SBs.
We demostrate that spicules naturally meet this condition because they are originating from filigrees in honeycomb like structures in the scale of granules lined up along supergranulation boundaries.  From the time-distance map of  spicules we derived a number distribution of inter-distances among spicules as wide as the waiting-time distribution of SBs \citep{DudokdeWit2020}. 

\end{enumerate}
These results suggest that small-scale solar ejections, although hardly traceable with current instruments, may retain a longer memory in their spatial distribution as earlier perceived by \cite{DudokdeWit2020}. In this sense the spatio-temporal distributions of spicule and associted filigrees manifest solar origin of the fine temporal structure of SBs.

We also addressed other issues which were once considered as shortcomings of spicules as an SB candidate: 

\begin{enumerate}

\item 
Although filigrees appear at the bases of spicules, they do not immediately respond to spicules, and threrefore are not JBPs of spicules. Instead they work as a placeholder for ejection, and the open fluxtubes rooted in the filigrees work as a conduit for spicules.  
The idea of JBP formation assumes reconnection occurring below the erupting materials, a.k.a. inner reconnection \citep{Moore2015,Sterling_2016b, Sterling_2020a}. If the reconnection occurs higher to bring up materials from below via rarefaction waves \citep{Scott_2022}, JBPs would hardly form underneath the cool ejecta.


\item 
As to the poor correlation between {\ha} spicules with EUV brightness, our sensitive magnetogram detected quasi-periodic emergence of the minor polarity patches in the region of the EUV intensity flickering and spicular activity. In other regions without EUV brightness, we fail to detect such a periodic magnetic activity. 
Such
a selective appearance of the EUV counterparts in regions with magnetic activity may imply the presence of a threshold for the correlation between spicule activity and EUV activity \citep[cf.][]{Nived_2022}. 

\item
We found temporal variations of spicules, filigrees, and magnetic parasitic polarity in the scale of 2, 6, and 10 minutes similar to the quasi-periods 3, 5, 10, and 20 minutes of the microstream-associated EUV fluctuations \citep{Kumar_2023b}. Our time scales, however, refer to duration of individual events occurring elsewhere rather than period of sustaining oscillations in fixed locations. 
We suggest that individual spicules occurring at multiple locations in the chromosphere might work as a periodic forcing to the corona where
more coherent oscillations or waves are generated as a consequence. 

  
\end{enumerate}

These conclusions lead us to a picture of sun-originated SBs largely different from other solar models for SBs. Instead of relying on specific magnetic eruption mechanisms known from larger events, 
our scenario suggests a more direct link between solar convection to small-scale solar wind transients via copious spicules.
Unlike other sporadic solar eruptions, spicules exhibit almost steady occurrence so that PSP may detect SBs at sufficiently high rate whenever they encounter open fields from CHBs.

\acknowledgements
 This work was supported by NSF grants, AGS-2114201 and AGS-2229064. It is also supported by NASA grants, 80NSSC19K0257 and 80NSSC20K1282.  BBSO/GST operation is supported by New Jersey Institute of Technology and US NSF grants AGS-1821294 and AGS-2309939. GST operation is partly supported by the Korea Astronomy and Space Science Institute and the Seoul National University. 
This paper is part of the Parker Solar Probe Focus entitled {\it Parker Solar Probe: Insights into the Physics of the Near-Solar Environment.}

\end{document}